\documentclass[aps,twocolumn,pra,showpacs,amsmath,amssymb,showkeys,10pt]{revtex4-2}
\usepackage{graphicx}
\usepackage{epstopdf}
\pdfoutput=1
\usepackage{braket}
\usepackage{hyperref}
\usepackage{longtable}

\allowdisplaybreaks

\begin{document}

	\title{Quantum electrodynamic description of ionization of the neutral hydrogen molecule}
	
	\author{Hui-hui Miao}
	\email[Correspondence to: Vorobyovy Gory 1, Moscow, 119991, Russia. E-mail address: ]{hhmiao@cs.msu.ru}
	\affiliation{Faculty of Computational Mathematics and Cybernetics, Lomonosov Moscow State University, Vorobyovy Gory 1, Moscow, Russia}

	\date{\today}

	\begin{abstract}	
	We investigate hydrogen molecule ionization within a unified framework combining finite-dimensional quantum electrodynamics with the Lindblad master equation, enabling systematic comparison across closed, dissipative, and influx-driven open systems. Our results reveal a universal tendency toward neutral $\mathrm{H}_2$ formation. Photon dissipation ($\gamma_\Omega$) accelerates stabilization, while electron ($\gamma_\mathrm{e}$) and phonon ($\gamma_\omega$) dissipation play distinct regulatory roles. Particle influx ($\mu_k$) induces complex energy redistribution, populating the atomic state $|\mathrm{H},\mathrm{H}\rangle$. The ionization pathway is highly sensitive to initial photon number and composition, which control spin-selective excitation channels. An embedded anode model confirms that orbital hybridization fundamentally constrains the maximum ionization probability to $3/4$. This work provides a unified theoretical foundation for quantum-controlled chemistry and cavity QED experiments.
	\end{abstract}
	
	\pacs{33.80.Eh, 42.50.Pq, 03.65.Yz, 31.15.ag}
	\keywords{hydrogen molecule ionization, finite-dimensional QED, Lindblad master equation, electron}

	\maketitle

\section{Introduction}
\label{sec:Intro}

The quantum dynamics of chemical processes \cite{Wang2021, McClean2021, Claudino2023}, particularly those involving light--matter interactions, represent a frontier in modern physical chemistry. The modeling of chemical reactions related to hydrogen is one of the main objectives of chemical modeling. Numerous theoretical articles \cite{Zhu2020, Ozhigov2021, ChenYou2022, Miao2023, MiaoOzhigov2023, You2025} have recently stimulated interest in this area, namely the hydroxyl cation ($\mathrm{OH}^+$) in a thermal bath \cite{ChenYou2022}, the neutral hydrogen molecule ($\mathrm{H}_2$) \cite{Miao2023, MiaoOzhigov2023}, and the hydrogen bond ($\mathrm{O}\cdots\mathrm{H}$) \cite{You2025}. Among these, the ionization of the hydrogen molecule stands as a fundamental prototype for understanding electron transfer and bond breaking. While the static electronic structure of $\mathrm{H}_2$ is well-understood, its real-time dynamics under driven and dissipative environments poses significant theoretical challenges. A comprehensive description requires a framework that can simultaneously account for the quantum nature of light, the internal degrees of freedom of the molecule, and the irreversible energy exchange with its surroundings.
	
	Toward this goal, models rooted in quantum electrodynamics (QED) are essential. The ultrastrong-coupling regime occurs when the coupling strength $g$ becomes comparable to the atomic ($\omega_\mathrm{a}$) or cavity ($\omega_\mathrm{c}$) frequencies, i.e., $\eta = \max(g/\hbar\omega_\mathrm{c}, g/\hbar\omega_\mathrm{a}) \in [0.1, 1)$ \cite{Haroche2013, Gu2017, Kockum2019, KockumMiranowicz2019, Forn-Diaz2019}. The quantum Rabi model \cite{Rabi1936, Rabi1937} is the fundamental description of the ultrastrong-coupling regime for a single two-level atom in a single-mode cavity, with the Dicke \cite{Dicke1954} and Hopfield \cite{Hopfield1958} models as its multi-atom and multi-mode generalizations. The regime $\eta \ge 1$ is known as deep strong coupling \cite{Casanova2010}. For $\eta < 0.1$, the simpler Jaynes--Cummings model \cite{Jaynes1963} applies, which describes a two-level atom interacting with a single-mode cavity field. Its multi-atom generalization is the Tavis--Cummings model \cite{Tavis1968}. Extending these to coupled cavities yields the Jaynes--Cummings--Hubbard model and Tavis--Cummings--Hubbard model \cite{Angelakis2007}. Many recent studies have advanced the field of QED models, exploring phase transitions \cite{Prasad2018, Wei2021}, quantum many-body phenomena \cite{OzhigovYI2020, Smith2021}, quantum gate implementations \cite{Dull2021}, and quantum correlations \cite{Miao2024, MiaoLi2025}. Further developments include supercomputing realizations of quantum modeling \cite{MiaoOzhigov2024, LiMiao2024}. The QED framework allows for a first-principles description of the interaction between the molecular electrons and the photon field, enabling the precise modeling of phenomena such as stimulated absorption and emission. However, to realistically describe open quantum systems --- where dissipation and decoherence are inevitable --- the QED Hamiltonian must be coupled with a master equation that governs the system's non-unitary evolution. The Lindblad master equation \cite{Alicki1979, Breuer2002, Kosloff2013} has emerged as the standard tool for this purpose, providing a mathematically rigorous method to model Markovian dynamics while preserving the positivity of the density matrix $\hat{\rho}$.

	In this work, we employ a tailored QED model described by the Lindblad master equation to investigate the ionization dynamics of a hydrogen molecule. This model is an extension of our previous association--dissociation model of the neutral hydrogen molecule \cite{Miao2023, MiaoOzhigov2023}. Our study systematically explores the system's behavior across three distinct regimes: the coherent, unitary evolution of a closed system; the dissipative dynamics of an open system with particle loss; and the more complex scenario involving the influx of particles. We meticulously analyze the roles of different dissipation channels --- for photons, electrons, and phonons --- and their impact on the formation of the hydrogen molecular cation ($\mathrm{H}_2^+$). Furthermore, we extend our model to incorporate a built-in anode, providing a novel perspective on the competition between electron escape and molecular stabilization. Our results delineate the system's inherent propensity to form $\mathrm{H}_2$, identify key dissipation parameters that either promote or suppress ionization, and reveal how initial quantum state preparation dictates the final outcome. By bridging coherent dynamics, dissipative processes, and external driving, this work offers a unified picture of hydrogen molecule ionization, providing insights that could inform experiments in quantum optics and the control of chemical reactions at the quantum level. A comprehensive list of symbols, abbreviations, and notation used throughout this paper is provided in the Appx. \ref{appxsec:List}.

	\section{Theoretical model}
	\label{sec:Model}
	
	\begin{figure*}[htbp]
	\centerline{\includegraphics[width=1.\textwidth]{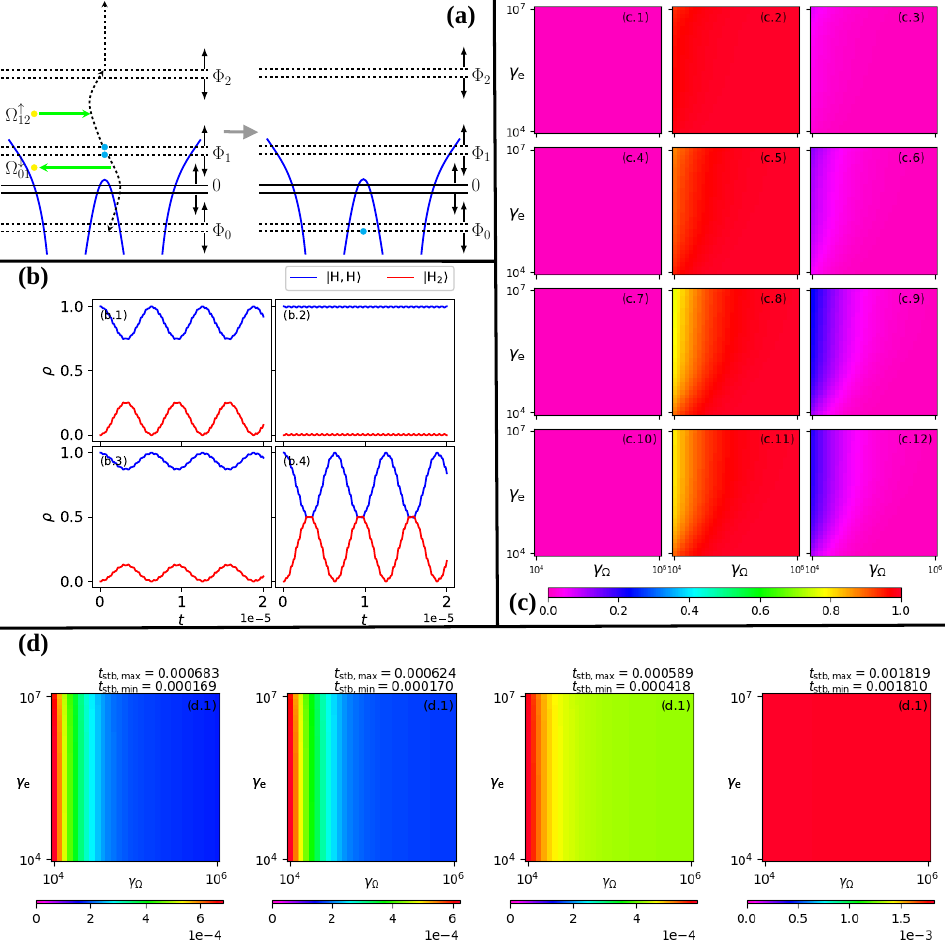}}
	\vspace*{8pt}
		\caption{(Color online) Ionization dynamics in a hydrogen molecule: unitary vs. dissipative evolution. Panel (a) depicts the ionization process of a hydrogen molecule. Here, the blue and yellow dots stand for electrons and photons, respectively. Panel (b) shows the unitary evolution under different initial conditions: (b.1) for initial states $|\Psi_\mathrm{initial}^0\rangle$, $|\Psi_\mathrm{initial}^1\rangle$, $|\Psi_\mathrm{initial}^2\rangle$, $|\Psi_\mathrm{initial}^3\rangle$, and $|\Psi_\mathrm{initial}^4\rangle$; (b.2) for $|\Psi_\mathrm{initial}^5\rangle$; (b.3) for $|\Psi_\mathrm{initial}^6\rangle$; (b.4) for $|\Psi_\mathrm{initial}^7\rangle$. Panel (c) shows the effect of different particle dissipation intensities on the ionization model (initial state is $|\Psi_\mathrm{initial}^6\rangle$): the horizontal and vertical axes of each subpanel represent the $\gamma_{\Omega}$ and $\gamma_\mathrm{e}$, respectively; the first through fourth rows correspond to $\gamma_{\omega}$ of $10^7$, $10^{7.5}$, $10^8$, and $10^{8.5}$, respectively; the first through third columns correspond to probabilities of $|\mathrm{H},\mathrm{H}\rangle$, $|\mathrm{H}_2\rangle$, and $|\mathrm{H}_2^+\rangle$, respectively. Here, the color bar for each heatmap ranges from 0 to 1. Panel (d) compares the time it takes for the system to reach a stable state: the four subpanels correspond to $\gamma_{\omega}$ of $10^7$, $10^{7.5}$, $10^8$, and $10^{8.5}$, respectively. Here, the color bar for each heatmap is scaled from 0 to the maximum value.}
		\label{fig:Model}
	\end{figure*}
	
	In our model, each energy level, whether atomic or molecular, is divided into two states: spin-up ($\uparrow$) and spin-down ($\downarrow$). According to the Pauli exclusion principle, each energy level may contain no more than one electron \cite{Pauli1925}. The schematic diagram of the target model is shown in Fig. \ref{fig:Model} (a), where the hybridization of atomic orbitals ($0_1$ and $0_2$) from two hydrogen atoms results in the formation of two molecular orbitals: a bonding orbital ($\Phi_0$) and an antibonding orbital ($\Phi_1$). These two molecular orbitals take the form:
	\begin{subequations}
		\label{eq:MolOrbitals}
		\begin{align}
			\Phi_0=\frac{1}{\sqrt{2}}\left(0_1+0_2\right),\label{eq:Phi0}\\
			\Phi_1=\frac{1}{\sqrt{2}}\left(0_1-0_2\right).\label{eq:Phi1}
		\end{align}
	\end{subequations}
	According to Eq. \eqref{eq:MolOrbitals}, we can obtain:
	\begin{subequations}
		\label{eq:AtOrbitals}
		\begin{align}
			0_1=\frac{1}{\sqrt{2}}\left(\Phi_0+\Phi_1\right),\label{eq:0_1}\\
			0_2=\frac{1}{\sqrt{2}}\left(\Phi_0-\Phi_1\right).\label{eq:0_2}
		\end{align}
	\end{subequations}
	After the molecular orbital is formed, two electrons with opposite spins, originally belonging to different atoms, occupy these two orbitals and undergo transitions between them through excitation--relaxation processes. Simultaneously, a transitional orbital ($\Phi_2$) exists farther away from the two atomic nuclei, allowing electrons to transition between orbitals $\Phi_2$ and $\Phi_1$ via excitation--relaxation. Taking an electron with $\uparrow$ in the antibonding orbital as an example, the electron absorbs a photon of mode $\Omega_{12}^{\uparrow}$ and transitions to the transitional orbital. Given its high energy and increased distance from the nuclei, the electron in the transitional state can be approximated as a free electron that has completely escaped the molecule. An analogous mechanism applies to the spin-down electron.  We stipulate that at most one electron can transition to the $\Phi_2$ orbital at any given time.

	The Hilbert space of quantum states of the entire system has the following form:
	\begin{equation}
		\label{eq:Space}
		|\Psi\rangle=|p_1\rangle_{\Omega_{12}^{\uparrow}}|p_2\rangle_{\Omega_{12}^{\downarrow}}|p_3\rangle_{\Omega_{01}^{\uparrow}}|p_4\rangle_{\Omega_{01}^{\downarrow}}|p_5\rangle_{\omega}|L\rangle_\mathrm{b}|k\rangle_\mathrm{n}|l_1\rangle_{\mathrm{e}^{\uparrow}}|l_2\rangle_{\mathrm{e}^{\downarrow}},
	\end{equation}
	where $p_1,\ p_2,\ p_3,\ p_4,\ p_5$ are the numbers of photons of modes $\Omega_{12}^{\uparrow}$, $\Omega_{12}^{\downarrow}$, $\Omega_{01}^{\uparrow}$, $\Omega_{01}^{\downarrow}$, $\omega$, respectively. Here, $\Omega_{12}^{\uparrow}$ is the mode of the photon required for an electron with $\uparrow$ to transition from orbital $\Phi_1$ to orbital $\Phi_2$, $\Omega_{12}^{\downarrow}$ is the mode of the photon required for an electron with $\downarrow$ to transition from orbital $\Phi_1$ to orbital $\Phi_2$, $\Omega_{01}^{\uparrow}$ is the mode of the photon required for an electron with $\uparrow$ to transition from orbital $\Phi_0$ to orbital $\Phi_1$, $\Omega_{01}^{\downarrow}$ is the mode of the photon required for an electron with $\downarrow$ to transition from orbital $\Phi_0$ to orbital $\Phi_1$. The state of the covalent bond is denoted by $|L\rangle_\mathrm{b}$: $L=0$ --- covalent bond formation, $L=1$ --- covalent bond breaking. The state of the nuclei is denoted by $|k\rangle_\mathrm{n}$: $k=0$ --- the nuclei are gathered together in one cavity; $k=1$ --- the nuclei are scattered in different cavities. $l_1,\ l_2$ describe the state of electrons: $l_1(l_2)=0$ --- electron with spin $\uparrow(\downarrow)$ in the orbital $\Phi_0$, $l_1(l_2)=1$ --- electron with spin $\uparrow(\downarrow)$ in the orbital $\Phi_1$, $l_1(l_2)=2$ --- electron with spin $\uparrow(\downarrow)$ in the orbital $\Phi_2$, $l_1(l_2)=\varnothing$ --- electron with spin $\uparrow(\downarrow)$ detaches from the transitional orbital and becomes a free electron.
	
	The dynamics of the system is described by solving the quantum master equation (QME) in the Markovian approximation for the density operator $\hat{\rho}$ of the system:
	\begin{equation}
		\label{eq:QME}
		\begin{aligned}
			i\hbar\dot{\hat{\rho}}&=\left[\hat{H},\hat{\rho}\right]+i\hat{L}\left(\hat{\rho}\right)\\
			&=\left[\hat{H},\hat{\rho}\right]+i\left[\sum_{k\in \mathcal{K}}\hat{L}_k\left(\hat{\rho}\right)+\sum_{k'\in \mathcal{K}'}\hat{L}_{k'}\left(\hat{\rho}\right)\right]\\
			&=\left[\hat{H},\hat{\rho}\right]+i\left[\sum_{k\in \mathcal{K}} \gamma_k\left(\hat{A}_k\hat{\rho}\hat{A}_k^{\dag}-\frac{1}{2}\left\{\hat{\rho},\hat{A}_k^{\dag}\hat{A}_k\right\}\right)\right.\\
			&\left.+\sum_{k'\in \mathcal{K}'}\gamma_{k'}\left(\hat{A}_k^{\dag}\hat{\rho}\hat{A}_k-\frac{1}{2}\left\{\hat{\rho},\hat{A}_k\hat{A}_k^{\dag}\right\}\right)\right],
		\end{aligned}
	\end{equation}
	where the term $\gamma_k$ refers to the overall spontaneous emission rate for photons for $k\in \mathcal{K}$. The total spontaneous influx rate for photons for $k'\in \mathcal{K}'$ is denoted by $\gamma_{k'}$. $\left[\hat{H},\hat{\rho}\right]=\hat{H}\hat{\rho}-\hat{\rho} \hat{H}$, $\left\{\hat{\rho},\hat{A}_k^{\dag}\hat{A}_k\right\}=\hat{\rho}\hat{A}_k^{\dag}\hat{A}_k+\hat{A}_k^{\dag}\hat{A}_k\hat{\rho}$, $\left\{\hat{\rho},\hat{A}_k\hat{A}_k^{\dag}\right\}=\hat{\rho}\hat{A}_k\hat{A}_k^{\dag}+\hat{A}_k\hat{A}_k^{\dag}\hat{\rho}$.
	
	The Hamiltonian of the model with consideration of the rotating wave approximation (RWA) \cite{Wu2007} has the following form
	\begin{equation}
		\label{eq:Hamil}
		\hat{H}=\hat{H}_\mathrm{atom}+\hat{H}_\mathrm{field}+\hat{H}_\mathrm{int}+\hat{H}_\mathrm{bond}+\hat{H}_\mathrm{tun},
	\end{equation}
	where $\hat{H}_\mathrm{atom}$ illustrates the atomic energy, $\hat{H}_\mathrm{field}$ shows the cavity field energy, $\hat{H}_\mathrm{int}$ depicts the interaction, $\hat{H}_\mathrm{bond}$ represents the formation of the covalent bond, $\hat{H}_\mathrm{tun}$ represents the tunneling effect.
	\begin{subequations}
		\label{eq:SubHamils}
		\begin{align}
			\hat{H}_\mathrm{atom}&=\hbar\Omega_{01}^{\uparrow}\hat{\sigma}_{\Omega_{01}^{\uparrow}}^{\dag}\hat{\sigma}_{\Omega_{01}^{\uparrow}}+\hbar\Omega_{01}^{\downarrow}\hat{\sigma}_{\Omega_{01}^{\downarrow}}^{\dag}\hat{\sigma}_{\Omega_{01}^{\downarrow}}\nonumber\\
			&+\hbar\left(\Omega_{01}^{\uparrow}+\Omega_{12}^{\uparrow}\right)\hat{\sigma}_{\Omega_{12}^{\uparrow}}^{\dag}\hat{\sigma}_{\Omega_{12}^{\uparrow}}\nonumber\\
			&+\hbar\left(\Omega_{01}^{\uparrow}+\Omega_{12}^{\uparrow}\right)\hat{\sigma}_{\Omega_{12}^{\downarrow}}^{\dag}\hat{\sigma}_{\Omega_{12}^{\downarrow}},\label{eq:HamilAtom}\\
			\hat{H}_\mathrm{field}&=\hbar\Omega_{01}^{\uparrow}\hat{a}_{\Omega_{01}^{\uparrow}}^{\dag}\hat{a}_{\Omega_{01}^{\uparrow}}+\hbar\Omega_{01}^{\downarrow}\hat{a}_{\Omega_{01}^{\downarrow}}^{\dag}\hat{a}_{\Omega_{01}^{\downarrow}}\nonumber\\
			&+\hbar\Omega_{12}^{\uparrow}\hat{a}_{\Omega_{12}^{\uparrow}}^{\dag}\hat{a}_{\Omega_{12}^{\uparrow}}+\hbar\Omega_{12}^{\downarrow}\hat{a}_{\Omega_{12}^{\downarrow}}^{\dag}\hat{a}_{\Omega_{12}^{\downarrow}},\label{eq:HamilField}\\
			\hat{H}_\mathrm{int}&=\left[g_{\Omega_{01}^{\uparrow}}\left(\hat{a}_{\Omega_{01}^{\uparrow}}^{\dag}\hat{\sigma}_{\Omega_{01}^{\uparrow}}+\hat{a}_{\Omega_{01}^{\uparrow}}\hat{\sigma}_{\Omega_{01}^{\uparrow}}^{\dag}\right)\right.\nonumber\\
			&+g_{\Omega_{01}^{\downarrow}}\left(\hat{a}_{\Omega_{01}^{\downarrow}}^{\dag}\hat{\sigma}_{\Omega_{01}^{\downarrow}}+\hat{a}_{\Omega_{01}^{\downarrow}}\hat{\sigma}_{\Omega_{01}^{\downarrow}}^{\dag}\right)\nonumber\\
			&+g_{\Omega_{12}^{\uparrow}}\left(\hat{a}_{\Omega_{12}^{\uparrow}}^{\dag}\hat{\sigma}_{\Omega_{12}^{\uparrow}}+\hat{a}_{\Omega_{12}^{\uparrow}}\hat{\sigma}_{\Omega_{12}^{\uparrow}}^{\dag}\right)\nonumber\\
			&\left.+g_{\Omega_{12}^{\downarrow}}\left(\hat{a}_{\Omega_{12}^{\downarrow}}^{\dag}\hat{\sigma}_{\Omega_{12}^{\downarrow}}+\hat{a}_{\Omega_{12}^{\downarrow}}\hat{\sigma}_{\Omega_{12}^{\downarrow}}^{\dag}\right)\right]\hat{\sigma}_{\omega}\hat{\sigma}_{\omega}^{\dag},\label{eq:HamilInt}\\
			\hat{H}_\mathrm{bond}&=\hbar\omega\hat{a}_{\omega}^{\dag}\hat{a}_{\omega}+\hbar\omega\hat{\sigma}_{\omega}^{\dag}\hat{\sigma}_{\omega}+g_{\omega}\left(\hat{a}_{\omega}^{\dag}\hat{\sigma}_{\omega}+\hat{a}_{\omega}\hat{\sigma}_{\omega}^{\dag}\right),\label{eq:HamilBond}\\
			\hat{H}_\mathrm{tun}&=\zeta\left(\hat{\sigma}_\mathrm{n}^{\dag}\hat{\sigma}_\mathrm{n}+\hat{\sigma}_\mathrm{n}\hat{\sigma}_\mathrm{n}^{\dag}\right)\hat{\sigma}_{\omega}^{\dag}\hat{\sigma}_{\omega},\label{eq:HamilTun}		
		\end{align}
	\end{subequations}
	where $\hbar=h/2\pi$ is the reduced Planck constant, $g$ is the coupling strength between the photon or phonon (with annihilation and creation operators $\hat{a}$ and $\hat{a}^{\dag}$, respectively) and the electron in the molecule (with excitation and relaxation operators $\hat{\sigma}^{\dag}$ and $\hat{\sigma}$, respectively): $g_{\Omega_{01}^{\uparrow}}$ is the coupling strength for the photon mode $\Omega_{01}^{\uparrow}$, $g_{\Omega_{01}^{\downarrow}}$ is the coupling strength for the photon mode $\Omega_{01}^{\downarrow}$, $g_{\Omega_{12}^{\uparrow}}$ is the coupling strength for the photon mode $\Omega_{12}^{\uparrow}$, $g_{\Omega_{12}^{\downarrow}}$ is the coupling strength for the photon mode $\Omega_{12}^{\downarrow}$, $g_{\omega}$ is the strength of formation or breaking of the covalent bond, $\zeta$ is the tunneling strength. $\hat{\sigma}_{\omega}\hat{\sigma}_{\omega}^{\dag}$ verifies that the covalent bond is formed, and $\hat{\sigma}_{\omega}^{\dag}\hat{\sigma}_{\omega}$ verifies that the covalent bond is broken.
	
	It is known that the ionization of a hydrogen molecule requires pumping photons with a mode of $\Omega_{12}$ ($\Omega_{12}^{\uparrow}$ or $\Omega_{12}^{\downarrow}$) into the system to promote electron transition to an unstable transitional orbital. Therefore, we postulate the following three initial conditions:
	\begin{itemize}
		\item The first condition is when the number of photons of mode $\Omega_{12}$ in the system is 1, i.e., $p_1+p_2=1$. This gives us three configurations:
		\begin{enumerate}
			\item $\{p_1=1,\ p_2=0\}$;
			\item $\{p_1=0,\ p_2=1\}$;
			\item Due to quantum superposition, we also consider their coherent and equally weighted superposition state: $\frac{1}{\sqrt{2}}\{p_1=1,\ p_2=0\}+\frac{1}{\sqrt{2}}\{p_1=0,\ p_2=1\}$.
		\end{enumerate}
		\item The second condition is when the number of photons of mode $\Omega_{12}$ in the system is 2, i.e., $p_1+p_2=2$. This gives us four configurations:
		\begin{enumerate}
			\item $\{p_1=2,\ p_2=0\}$;
			\item $\{p_1=0,\ p_2=2\}$;
			\item $\{p_1=1,\ p_2=1\}$;
			\item Due to quantum superposition, we also consider their coherent superposition state: $\frac{1}{2}\{p_1=2,\ p_2=0\}+\frac{1}{2}\{p_1=0,\ p_2=2\}+\frac{1}{\sqrt{2}}\{p_1=1,\ p_2=1\}$, which indicates that the probability of having two photons of same mode $\Omega_{12}^{\uparrow}$ (or $\Omega_{12}^{\downarrow}$) in the system is $\frac{1}{4}$, whereas the probability of having one photon of mode $\Omega_{12}^{\uparrow}$ and another of mode $\Omega_{12}^{\downarrow}$ is $\frac{1}{2}$.
		\end{enumerate}
		\item The third condition is when the number of photons of mode $\Omega_{12}$ in the system is 0, i.e., $\{p_1=0,\ p_2=0\}$.
	\end{itemize}
	Therefore, we obtain eight initial states:
	\begin{subequations}
		\label{eq:InitialState}
		\begin{align}
			|\Psi_\mathrm{initial}^0\rangle&=|1\rangle|0\rangle\otimes|\Psi_\mathrm{f}\rangle,\label{eq:Initial0}\\
			|\Psi_\mathrm{initial}^1\rangle&=|0\rangle|1\rangle\otimes|\Psi_\mathrm{f}\rangle,\label{eq:Initial1}\\
			|\Psi_\mathrm{initial}^2\rangle&=\left(\frac{1}{\sqrt{2}}|1\rangle|0\rangle+\frac{1}{\sqrt{2}}|0\rangle|1\rangle\right)\otimes|\Psi_\mathrm{f}\rangle,\label{eq:Initial2}\\
			|\Psi_\mathrm{initial}^3\rangle&=|2\rangle|0\rangle\otimes|\Psi_\mathrm{f}\rangle,\label{eq:Initial3}\\
			|\Psi_\mathrm{initial}^4\rangle&=|0\rangle|2\rangle\otimes|\Psi_\mathrm{f}\rangle,\label{eq:Initial4}\\
			|\Psi_\mathrm{initial}^5\rangle&=|1\rangle|1\rangle\otimes|\Psi_\mathrm{f}\rangle,\label{eq:Initial5}\\
			|\Psi_\mathrm{initial}^6\rangle&=\left(\frac{1}{2}|2\rangle|0\rangle+\frac{1}{2}|0\rangle|2\rangle+\frac{1}{\sqrt{2}}|1\rangle|1\rangle\right)\otimes|\Psi_\mathrm{f}\rangle,\label{eq:Initial6}\\
			|\Psi_\mathrm{initial}^7\rangle&=|0\rangle|0\rangle\otimes|\Psi_\mathrm{f}\rangle,\label{eq:Initial7}
		\end{align}
	\end{subequations}
	where the first two qubits correspond to $|p_1\rangle_{\Omega_{12}^{\uparrow}}|p_2\rangle_{\Omega_{12}^{\downarrow}}$ in Eq. \eqref{eq:Space}, while the remaining qubits, being fixed, are collectively denoted as $|\Psi_\mathrm{f}\rangle$, which is written as follows:
		\begin{equation}
			\label{eq:Initialfixed}
			|\Psi_\mathrm{f}\rangle=|0\rangle|0\rangle|0\rangle|1\rangle|1\rangle\otimes\frac{1}{2}\left(|0\rangle|0\rangle-|0\rangle|1\rangle+|1\rangle|0\rangle-|1\rangle|1\rangle\right),
		\end{equation}
	where the first three qubits $|0\rangle|0\rangle|0\rangle$ correspond to $|p_3\rangle_{\Omega_{01}^{\uparrow}}|p_4\rangle_{\Omega_{01}^{\downarrow}}|p_5\rangle_{\omega}$ in Eq. \eqref{eq:Space}, and show that the number of photons of mode $\Omega_{01}^{\uparrow}$, the number of photons of mode $\Omega_{01}^{\downarrow}$, and the number of phonons of mode $\omega$ all equal to 0; the fourth qubit $|1\rangle$ corresponds to $|L\rangle_\mathrm{b}$, and indicates that the covalent bond is broken; the fifth qubit $|1\rangle$ corresponds to $|k\rangle_\mathrm{n}$, and indicates the state of the nuclei, scattering in different cavities. We now focus on the last two qubits corresponding to the electronic states $|l_1\rangle_{\mathrm{e}^{\uparrow}}|l_2\rangle_{\mathrm{e}^{\downarrow}}$. Given that $L=1$ and $k=1$, the initial state is defined as having no covalent bond formed, and with the two hydrogen atoms located in separate cavities. Consequently, the electrons reside in their atomic orbitals. Assuming each atom carries one electron with opposite spins, we denote this electronic state as $|0_1\rangle|0_2\rangle$. According to Eq. \eqref{eq:AtOrbitals}, $|0_1\rangle|0_2\rangle$ can be written as:
	\begin{equation}
		\label{eq:InitialAtOrbital}
		\begin{aligned}
			|0_1\rangle|0_2\rangle&=\frac{1}{\sqrt{2}}\left(|\Phi_0\rangle+|\Phi_1\rangle\right)\otimes\frac{1}{\sqrt{2}}\left(|\Phi_0\rangle-|\Phi_1\rangle\right)\\
			&=\frac{1}{2}\left(|\Phi_0\rangle|\Phi_0\rangle-|\Phi_0\rangle|\Phi_1\rangle+|\Phi_1\rangle|\Phi_0\rangle-|\Phi_1\rangle|\Phi_1\rangle\right)\\
			&=\frac{1}{2}\left(|0\rangle|0\rangle-|0\rangle|1\rangle+|1\rangle|0\rangle-|1\rangle|1\rangle\right).
		\end{aligned}
	\end{equation}
	
	\section{Numerical results} 
	\label{sec:Results}
	
	In simulations: $\hbar=1$, $\Omega_{12}^{\uparrow}=\Omega_{12}^{\downarrow}=10^{10}$, $\Omega_{01}^{\uparrow}=\Omega_{01}^{\downarrow}=10^9$, $\omega=10^8$, $g_{\Omega_{12}^{\uparrow}}=g_{\Omega_{12}^{\downarrow}}=10^8$, $g_{\Omega_{01}^{\uparrow}}=g_{\Omega_{01}^{\downarrow}}=10^7$, $g_{\omega}=10^6$, and $\zeta=10^7$.
	
	\subsection{Numerical method} 
	\label{subsec:Method}

	The solution $\hat{\rho}\left(t\right)$ in Eq. \eqref{eq:QME} may be approximately found as a sequence of two steps: in the first step we take one step in solving the unitary part of Eq. \eqref{eq:QME}:
	\begin{equation}
		\label{eq:UnitaryPart}
		\tilde{\hat{\rho}}\left(t+\Delta t\right)=\exp\left({-\frac{i}{\hbar}\hat{H}\Delta t}\right)\hat{\rho}\left(t\right)\exp\left(\frac{i}{\hbar}\hat{H}\Delta t\right).
	\end{equation}
In the second step, we take one step in the solution of Eq. \eqref{eq:QME} with the commutator removed:
	\begin{equation}
		\label{eq:Solution}
		\hat{\rho}\left(t+\Delta t\right)=\tilde{\hat{\rho}}\left(t+\Delta t\right)+\frac{1}{\hbar}\hat{L}\left(\tilde{\hat{\rho}}(t+\Delta t)\right)\Delta t.
	\end{equation}
	
	Evidently, computing the matrix exponential in Eq. \eqref{eq:UnitaryPart} is a key step. While several classical mathematical methods have been proposed \cite{Moler1978, Moler2003} for this purpose, in this paper the precise time step integration method (PTSIM) \cite{Zhong1991, ZhongWilliams1994, Zhong1994, Zhong1995, GAO2016} of matrix exponential is introduced. When solving the equations of motion using stepwise integration, the matrix exponential operation involved is $e^{A\Delta t}$. According to the addition theorem of matrix exponential:
	\begin{equation}
		\label{eq:MatrixExp}
		e^{A\Delta t}=\left[e^{A\frac{\Delta t}{2^M}}\right]^{2^M}=\left[e^{A\varepsilon}\right]^{2^M},
	\end{equation}
	where $A$ is a matrix, $\Delta t$ is time step, $\varepsilon=\frac{\Delta t}{2^M}$ (usually $M$ is taken to be equal to 20). The matrix exponential can be approximated using Taylor series expansion (4 terms):
	\begin{equation}
		\label{eq:Taylor}
		e^{A\varepsilon}\approx I+A\varepsilon+\frac{(A\varepsilon)^2}{2!}+\frac{(A\varepsilon)^3}{3!}+\frac{(A\varepsilon)^4}{4!},
	\end{equation}
	where $I$ is the unit matrix. Then,
	\begin{equation}
		\label{eq:PTSIM}
		\begin{aligned}
			e^{A\Delta t}&\approx \left[I+A\varepsilon+\frac{(A\varepsilon)^2}{2!}+\frac{(A\varepsilon)^3}{3!}+\frac{(A\varepsilon)^4}{4!}\right]^{2^M}\\
			&=\left[I+T_{a,0}\right]^{2^M},
		\end{aligned}
	\end{equation}
	where $T_{a,0}=A\varepsilon+\frac{(A\varepsilon)^2}{2!}+\frac{(A\varepsilon)^3}{3!}+\frac{(A\varepsilon)^4}{4!}$. Then
	\begin{equation}
		\label{eq:Recursion}
		\begin{aligned}
			\left[I+T_{a,0}\right]^{2^M}&=\left[(I+T_{a,0})^2\right]^{2^{M-1}}\\
			&=\left[I+2T_{a,0}+T_{a,0}T_{a,0}\right]^{2^{M-1}}\\
			&=\left[I+T_{a,1}\right]^{2^{M-1}}\\
			&=\left[I+T_{a,2}\right]^{2^{M-2}}\\
			&\cdots\cdots\\
			&=\left[I+T_{a,M}\right],
		\end{aligned}
	\end{equation}
	where $T_{a,n}=2T_{a,n-1}+T_{a,n-1}T_{a,n-1},\ n\geq1$. The above calculation process avoids operations between values with large differences in magnitude and avoids the impact of rounding errors.
	
	\subsection{Subspaces} 
	\label{subsec:Subspaces}
	
	Prior to analyzing the results, we first establish the definitions of the three following subspaces:
	\begin{widetext}
	\begin{subequations}
    		\label{eq:Subspaces}
    		\begin{align}
        		|\mathrm{H},\mathrm{H}\rangle&=\mathrm{span}\left\{|\cdots\rangle|L\rangle_\mathrm{b}|\cdots\rangle\ \Big|\ L=1\right\}, \label{eq:SubspaceAts} \\
        		|\mathrm{H}_2\rangle&=\mathrm{span}\left\{|\cdots\rangle|L\rangle_\mathrm{b}|k\rangle_\mathrm{n}|l_1\rangle_{\mathrm{e}^\uparrow}|l_2\rangle_{\mathrm{e}^\downarrow}\ \Big|\ L=0,\ k=0,\ \mathrm{AND}(l_1\neq\varnothing,\ l_2\neq\varnothing)\right\}, \label{eq:SubspaceMol} \\
        		|\mathrm{H}_2^+\rangle&=\mathrm{span}\left\{|\cdots\rangle|L\rangle_\mathrm{b}|k\rangle_\mathrm{n}|l_1\rangle_{\mathrm{e}^\uparrow}|l_2\rangle_{\mathrm{e}^\downarrow}\ \Big|\ L=0,\ k=0,\ \mathrm{XOR}(l_1=\varnothing,\ l_2=\varnothing)\right\}, \label{eq:SubspaceIon}
    		\end{align}
	\end{subequations}
	\end{widetext}
	where qubits without conditional restrictions are denoted by $|\cdots\rangle$. We partition the total Hilbert space into three subspaces, each corresponding to a distinct system state:
	\begin{itemize}
		\item Subspace $|\mathrm{H},\mathrm{H}\rangle$: Two independent hydrogen atoms ($L=1$ signifies no covalent bond formation).
		\item Subspace $|\mathrm{H}_2\rangle$: A neutral hydrogen molecule ($L=0$ signifies covalent bond formation; $k=0$ signifies both atoms are in the same cavity. The $\mathrm{AND}$ logical operator requires both $l_1\neq\varnothing$ and $l_2\neq\varnothing$).
		\item Subspace $|\mathrm{H}_2^+\rangle$: A hydrogen molecular cation ($L=0$ signifies covalent bond formation; $k=0$ signifies both atoms are in the same cavity. The $\mathrm{XOR}$ logical operator requires that either $l_1=\varnothing$ or $l_2=\varnothing$, but not both).
	\end{itemize}
	In practice, as we employ the Lindblad master equation (Eqs. \eqref{eq:UnitaryPart} and \eqref{eq:Solution}) which acts on the density matrix $\hat{\rho}$, the diagonal elements of $\hat{\rho}$ represent probabilities. Thus, the probabilities for $|\mathrm{H},\mathrm{H}\rangle$, $|\mathrm{H}_2\rangle$, and $|\mathrm{H}_2^+\rangle$ are obtained by summing their respective associated diagonal elements.

	\subsection{Unitary evolution} 
	\label{subsec:Unitary}

	We first consider an ideal, closed system where no particle inflow or escape occurs, meaning parameters $\gamma_k$ and $\gamma_{k'}$ in Eq. \eqref{eq:QME} are set 0. Under this condition, the Lindblad master equation reduces to the von Neumann equation:
	\begin{equation}
		\label{eq:VNE}
		i\hbar\dot{\hat{\rho}}=\left[\hat{H},\hat{\rho}\right].
	\end{equation}
	
	In the closed system, the absence of electron escape restricts the dynamics to oscillations between $|\mathrm{H},\mathrm{H}\rangle$ and $|\mathrm{H}_2\rangle$. We investigated the influence of eight different initial states on these oscillations, with the results presented in Fig. \ref{fig:Model} (b). In subpanel (b.1), we observe that the oscillation curves for initial states $|\Psi_\mathrm{initial}^0\rangle\sim|\Psi_\mathrm{initial}^4\rangle$ completely coincide. This is because photons of mode $\Omega_{12}^{\uparrow}$ and $\Omega_{12}^{\downarrow}$ are energetically degenerate. Therefore, at the initial time and under otherwise identical conditions, pumping in one photon of mode $\Omega_{12}^{\uparrow}$ ($|\Psi_\mathrm{initial}^0\rangle$) or one photon of mode $\Omega_{12}^{\downarrow}$ ($|\Psi_\mathrm{initial}^1\rangle$) results in identical dynamics. Furthermore, the coherent, equal-weight superposition of states $|\Psi_\mathrm{initial}^0\rangle$ and $|\Psi_\mathrm{initial}^1\rangle$ ($|\Psi_\mathrm{initial}^2\rangle$) also produces the same curve. Since the system contains only one spin-up electron, the oscillation curve remains identical whenever the population of photons of mode $\Omega_{12}^{\uparrow}$ is greater than or equal to 1. Consequently, the curve for $|\Psi_\mathrm{initial}^3\rangle$ coincides with that of $|\Psi_\mathrm{initial}^0\rangle$, and by the same reasoning, the curve for $|\Psi_\mathrm{initial}^4\rangle$ coincides with that of $|\Psi_\mathrm{initial}^1\rangle$, too. For the initial state $|\Psi_\mathrm{initial}^5\rangle$, which contains a photon of mode $\Omega_{12}^{\uparrow}$ and a photon of mode $\Omega_{12}^{\downarrow}$, both electrons with opposite spins have the opportunity to be excited into orbital $\Phi_2$. Consequently, we obtain an oscillation curve distinct from those in (b.1), as shown in (b.2). Evidently, the superposition state of $|\Psi_\mathrm{initial}^3\rangle$, $|\Psi_\mathrm{initial}^4\rangle$, and $|\Psi_\mathrm{initial}^5\rangle$ ($|\Psi_\mathrm{initial}^6\rangle$) results from the combination of the two oscillation curves in subpanels (b.1) and (b.2), as shown in (b.3). Meanwhile, when there are neither photons of mode $\Omega_{12}^{\uparrow}$ nor photons of mode $\Omega_{12}^{\downarrow}$ initially ($|\Psi_\mathrm{initial}^7\rangle$), the system exhibits an oscillation curve with the maximum amplitude, as shown in (b.4). From subpanels (b.1) to (b.4), we find that in the closed system, the excitation of electrons between orbital $\Phi_1$ and orbital $\Phi_2$ significantly reduces the amplitude of the oscillation curves.
	
	\subsection{Dissipative evolution} 
	\label{subsec:Dissipative}
	
	We now investigate the dissipative dynamics in the open system and the influence of different dissipation intensities on the formation of the $\mathrm{H}_2^+$. We assume that two photons are pumped into the system randomly at the initial time, with equal probability for a photon of mode $\Omega_{12}^{\uparrow}$ and a photon of mode $\Omega_{12}^{\downarrow}$, resulting in the initial state $|\Psi_\mathrm{initial}^6\rangle$. Due to the dissipative processes, the oscillations are damped, and the system eventually stabilizes after a sufficiently long evolution. The results in panel (c) of Fig. \ref{fig:Model} show that the energy ultimately flows into two subspaces: $|\mathrm{H}_2\rangle$ and $|\mathrm{H}_2^+\rangle$. Furthermore, electron dissipation intensities ($\gamma_\mathrm{e}$) and phonon dissipation intensities ($\gamma_{\omega}$) promote the formation of the hydrogen molecular cation, whereas photon dissipation intensities ($\gamma_{\Omega}$) suppress it. Our results consistently show that the population of $|\mathrm{H}_2\rangle$ is significantly higher than that of $|\mathrm{H}_2^+\rangle$ under all conditions, indicating the system's inherent tendency to form a stable neutral hydrogen molecule. For the time-based dynamic changes of the heatmap in panel (c), please refer to the \href{https://drive.google.com/drive/folders/1hZpICSbDHC9GW6X_1iPWRwpPUGFOp8mZ?usp=sharing}{animation-1.mp4}.
	
\begin{figure*}[htbp]
	\centerline{\includegraphics[width=.9\textwidth]{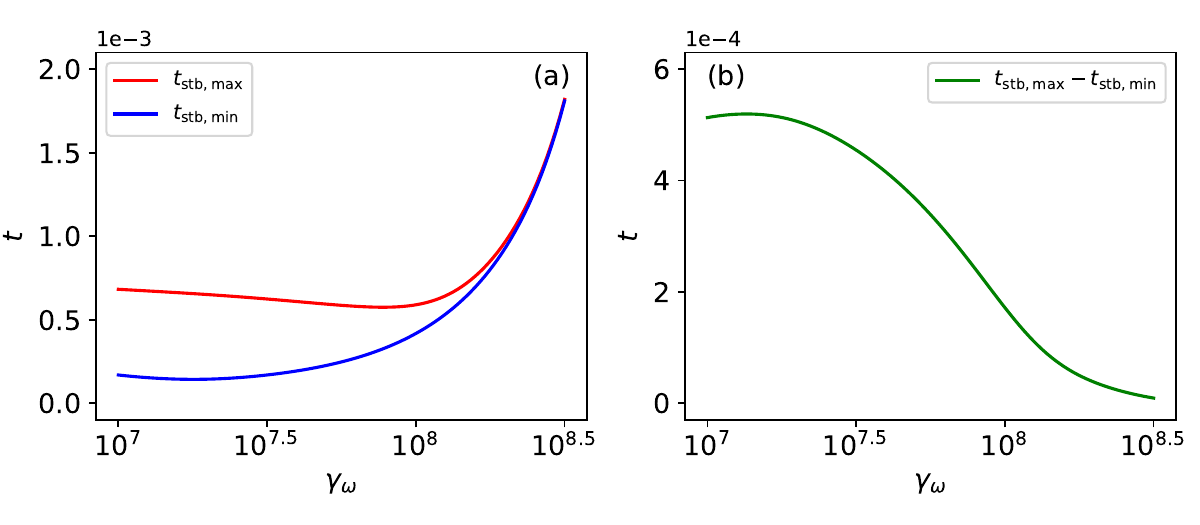}}
	\vspace*{8pt}
		\caption{(Color online) Comparison of the time it takes for the system to reach a stable state. As shown in Fig. \ref{fig:Model} (d), we investigated the impact of $\gamma_{\Omega}$ and $\gamma_\mathrm{e}$ on the system's stabilization time for $\gamma_{\omega}=10^7,\ 10^{7.5},\ 10^8,$ and $10^{8.5}$, identifying the maximum and minimum times for each case. We then analyzed how these extreme values evolve as $\gamma_{\omega}$ increases from $10^7$ to $10^{8.5}$ (panel (a)). Additionally, the difference between them is plotted against $\gamma_{\omega}$ in panel (b).}
		\label{fig:ComparisonMaxMin}
	\end{figure*}	
	
\begin{figure}[htbp]
	\centerline{\includegraphics[width=.5\textwidth]{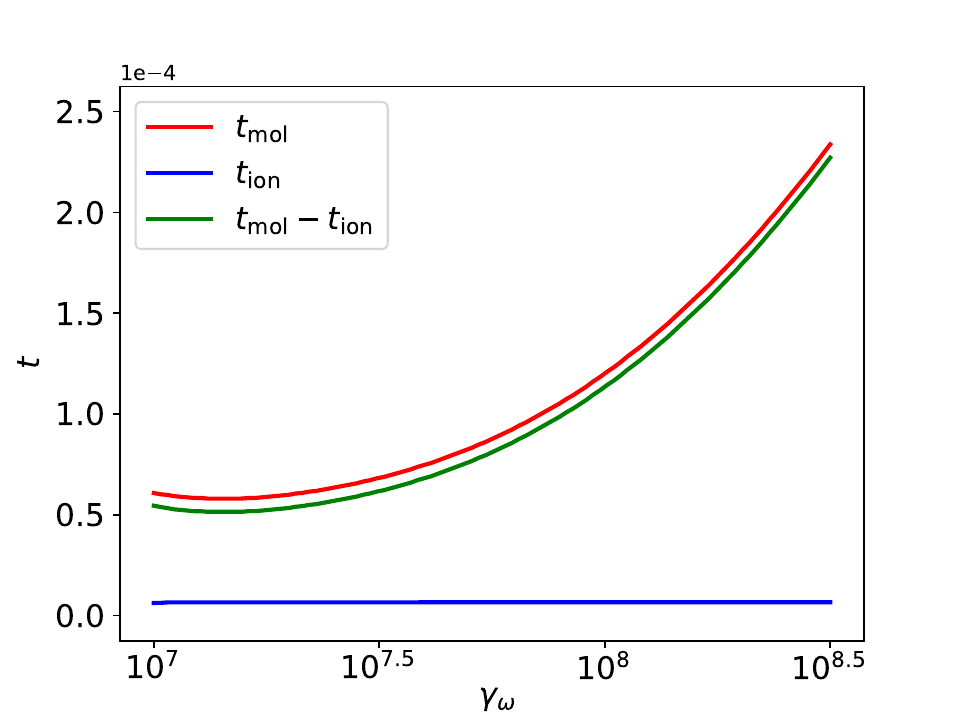}}
	\vspace*{8pt}
		\caption{(Color online) Comparison of the time required for hydrogen molecules and hydrogen molecular ions to reach stability. We consider one scenario with $\gamma_{\Omega}=10^6$ and $\gamma_\mathrm{e}=10^7$. We then analyze how these extreme values evolve as $\gamma_{\omega}$ increases from $10^7$ to $10^{8.5}$. The time for hydrogen molecules to reach stability (shown in red) and the time for hydrogen molecular ions to reach stability (shown in blue) are plotted as functions of $\gamma_{\omega}$, and their difference is shown in green.}
		\label{fig:ComparisonMolIon}
	\end{figure}	
	
\begin{figure*}[htbp]
	\centerline{\includegraphics[width=.9\textwidth]{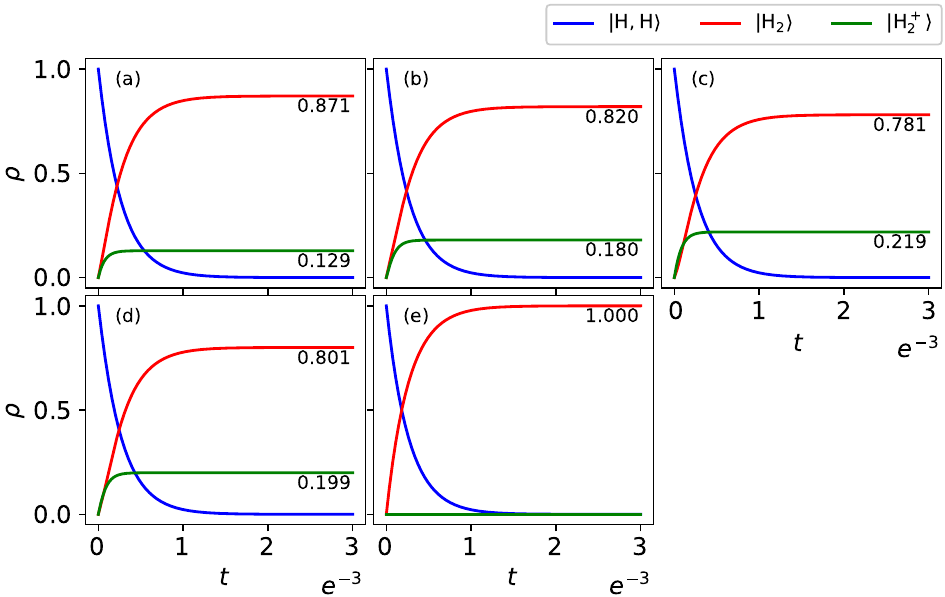}}
	\vspace*{8pt}
		\caption{(Color online) Dissipative dynamics for different initial conditions. Panel (a) shows the dynamics for $|\Psi_\mathrm{initial}^0\rangle$, $|\Psi_\mathrm{initial}^1\rangle$, and $|\Psi_\mathrm{initial}^2\rangle$; panel (b) for $|\Psi_\mathrm{initial}^3\rangle$ and $|\Psi_\mathrm{initial}^4\rangle$; panel (c) for $|\Psi_\mathrm{initial}^5\rangle$; panel (d) for $|\Psi_\mathrm{initial}^6\rangle$; and panel (e) for $|\Psi_\mathrm{initial}^7\rangle$.}
		\label{fig:Dissipation}
	\end{figure*}
	
	In panel (c) of Fig. \ref{fig:Model}, we investigate the influence of $\gamma_{\Omega}$ and $\gamma_\mathrm{e}$ on $|\mathrm{H}_2^+\rangle$ formation for $\gamma_{\omega}=10^7$, $10^{7.5}$, $10^8$, and $10^{8.5}$. We also determine the stabilization time $t_\mathrm{stb}$ for these four cases, defined as the time when the combined probability of $|\mathrm{H}_2\rangle$ and $|\mathrm{H}_2^+\rangle$ exceeds 0.999, and present the results as heatmaps in panel (d). The data reveal that $\gamma_{\Omega}$ significantly accelerates stabilization, whereas $\gamma_\mathrm{e}$ has a negligible effect. This is because the system evolves toward the stable hydrogen molecule, a process promoted by $\gamma_{\Omega}$. Consequently, a larger $\gamma_{\Omega}$ leads to faster stabilization. Furthermore, as $\gamma_{\omega}$ increases, the difference between the maximum value $t_\mathrm{stb,max}$ and minimum value $t_\mathrm{stb,min}$ in the heatmaps decreases. This indicates that the promoting effect of $\gamma_{\Omega}$ becomes less pronounced at high $\gamma_{\omega}$ values. Fig. \ref{fig:ComparisonMaxMin} plots $t_\mathrm{stb,max}$ and $t_\mathrm{stb,min}$ against $\gamma_{\omega}$. Panel (a) of Fig. \ref{fig:ComparisonMaxMin} shows that $t_\mathrm{stb,max}$ decreases slowly from $\gamma_{\omega}=10^7$ to $10^8$, beyond which it increases exponentially. In contrast, $t_\mathrm{stb,min}$ begins to rise when $\gamma_{\omega}<10^{7.5}$. As shown in panel (b) of Fig. \ref{fig:ComparisonMaxMin}, the difference between $t_\mathrm{stb,max}$ and $t_\mathrm{stb,min}$ rapidly decreases to 0 with increasing $\gamma_{\omega}$. This indicates that for $\gamma_{\omega}>10^8$, the influence of $\gamma_{\omega}$ on $t_\mathrm{stb}$ is far greater than that of $\gamma_{\Omega}$ and $\gamma_\mathrm{e}$. Additionally, we investigate the stabilization time of the neutral hydrogen molecule ($t_{\mathrm{mol}}$) and that of the hydrogen molecular cation ($t_{\mathrm{ion}}$) as functions of $\gamma_\omega$ (varying from $10^7$ to $10^{8.5}$), under fixed values of $\gamma_e = 10^7$ and $\gamma_\Omega = 10^6$. We assume that $t_{\mathrm{mol}}$ is obtained when $\left(\hat{\rho}_{|\mathrm{H}_2\rangle}(t+\Delta t)-\hat{\rho}_{|\mathrm{H}_2\rangle}(t)\right)/\hat{\rho}_{|\mathrm{H}_2\rangle}(t) < 10^{-5}$, and similarly, $t_{\mathrm{ion}}$ is obtained when $\left(\hat{\rho}_{|\mathrm{H}_2^+\rangle}(t+\Delta t)-\hat{\rho}_{|\mathrm{H}_2^+\rangle}(t)\right)/\hat{\rho}_{|\mathrm{H}_2^+\rangle}(t) < 10^{-5}$. Here, we choose $10^{-5}$ rather than a smaller value to prevent numerical noise from affecting the results. As shown in Fig. \ref{fig:ComparisonMolIon}, $t_{\mathrm{ion}}$ is much smaller than $t_{\mathrm{mol}}$. Moreover, $t_{\mathrm{ion}}$ is largely insensitive to the variation of $\gamma_\omega$, while $t_{\mathrm{mol}}$ increases with increasing $\gamma_\omega$.
	
	As can be seen from Fig. \ref{fig:Model} (c), the probability of obtaining the hydrogen molecular cation reaches its maximum when $\gamma_{\Omega}$ is at its minimum ($\gamma_{\Omega}^\mathrm{min}=10^4$) while $\gamma_{\omega}$ and $\gamma_\mathrm{e}$ are at their maxima ($\gamma_{\omega}^\mathrm{max}=10^{8.5}$, $\gamma_\mathrm{e}^\mathrm{max}=10^7$). Under these optimized conditions, we simulate the dissipative evolution of the system for different initial states, as shown in Fig. \ref{fig:Dissipation}. In each panel, due to the presence of dissipation, the probability of $|\mathrm{H},\mathrm{H}\rangle$ gradually decays from 1 to 0. In the closed system, the unitary evolution from $|\Psi_\mathrm{initial}^0\rangle$ to $|\Psi_\mathrm{initial}^4\rangle$ is identical (see Fig. \ref{fig:Model} (b)). In contrast, under dissipative evolution, the dynamics separate into two distinct cases, shown in Fig. \ref{fig:Dissipation} (a) and (b). Evidently, under dissipative evolution, the probability of obtaining the $\mathrm{H}_2^+$ is higher for initial states $|\Psi_\mathrm{initial}^3\rangle$ and $|\Psi_\mathrm{initial}^4\rangle$ than for $|\Psi_\mathrm{initial}^0\rangle$, $|\Psi_\mathrm{initial}^1\rangle$, and $|\Psi_\mathrm{initial}^2\rangle$. This is attributed to the fact that $|\Psi_\mathrm{initial}^3\rangle$ ($|\Psi_\mathrm{initial}^4\rangle$) contains two free photons of mode $\Omega_{12}^{\uparrow}$ ($\Omega_{12}^{\downarrow}$), whereas $|\Psi_\mathrm{initial}^0\rangle$, $|\Psi_\mathrm{initial}^1\rangle$, and $|\Psi_\mathrm{initial}^2\rangle$ contain only one free photon of $\Omega_{12}$. A greater population of these specific photons more effectively promotes the upward transition of electrons to the $\Phi_2$ orbital, thereby facilitating the ionization of the hydrogen molecule. A comparison between panels (b) and (c) indicates that, for the same initial total photon number of two, an initial state comprising two photons of different modes ($\Omega_{12}^{\uparrow}$ and $\Omega_{12}^{\downarrow}$) is more effective at promoting hydrogen molecule ionization than a state with two photons of a single mode. This enhanced efficiency arises because state $|\Psi_\mathrm{initial}^5\rangle$ can excite either a spin-up or a spin-down electron to the $\Phi_2$ orbital. In contrast, $|\Psi_\mathrm{initial}^3\rangle$ can only excite a spin-up electron, and $|\Psi_\mathrm{initial}^4\rangle$ can only excite a spin-down electron. Furthermore, the characteristics of the annihilation operator used in the model constitute a second contributing factor to this observed difference. The matrix forms of the annihilation and creation operators are given by:
	\begin{subequations}
		\label{eq:OperatorsA}
		\begin{align}
			\hat{a}&=\begin{array}{c@{\hspace{-5pt}}l}
			 \begin{array}{c}
			 	|0\rangle \\
			 	|1\rangle \\
			 	|2\rangle \\
			 	\vdots \\
			 	\vdots \\
			 	|p-2\rangle \\
			 	|p-1\rangle \\
			 	|p\rangle \\
			 \end{array}
			 & \left(
			 \begin{array}{cccccccc}
			 	0 & 1 & 0 & \cdots & \cdots & 0 & 0 & 0 \\
			 	0 & 0 & \sqrt{2} & \cdots & \cdots & 0 & 0 & 0 \\
			 	0 & 0 & 0 & \cdots & \cdots & 0 & 0 & 0 \\
			 	\vdots & \vdots & \vdots & \ddots & \ddots & \vdots & \vdots & \vdots \\
			 	\vdots & \vdots & \vdots & \ddots & \ddots & \vdots & \vdots & \vdots \\
			 	0 & 0 & 0 & \cdots & \cdots & 0 & \sqrt{p-1} & 0 \\
			 	0 & 0 & 0 & \cdots & \cdots & 0 & 0 & \sqrt{p}\\
			 	0 & 0 & 0 & \cdots & \cdots & 0 & 0 & 0 \\
			 \end{array}
			 \right)
			\end{array},\label{eq:Annihilation}\\
			\hat{a}^{\dag}&=\begin{array}{c@{\hspace{-5pt}}l}
			 \begin{array}{c}
			 	|0\rangle \\
			 	|1\rangle \\
			 	|2\rangle \\
			 	\vdots \\
			 	\vdots \\
			 	|p-2\rangle \\
			 	|p-1\rangle \\
			 	|p\rangle \\
			 \end{array}
			 & \left(
			 \begin{array}{cccccccc}
			 	0 & 0 & 0 & \cdots & \cdots & 0 & 0 & 0 \\
			 	1 & 0 & 0 & \cdots & \cdots & 0 & 0 & 0 \\
			 	0 & \sqrt{2} & 0 & \cdots & \cdots & 0 & 0 & 0 \\
			 	\vdots & \vdots & \vdots & \ddots & \ddots & \vdots & \vdots & \vdots \\
			 	\vdots & \vdots & \vdots & \ddots & \ddots & \vdots & \vdots & \vdots \\
			 	0 & 0 & 0 & \cdots & \cdots & 0 & 0 & 0 \\
			 	0 & 0 & 0 & \cdots & \cdots & \sqrt{p-1} & 0 & 0 \\
			 	0 & 0 & 0 & \cdots & \cdots & 0 & \sqrt{p} & 0 \\
			 \end{array}
			 \right)
			\end{array}.\label{eq:Creation}
		\end{align}
	\end{subequations}
	According to the aforementioned operators, when the particle number increases from $p-1$ to $p$ or decreases from $p$ to $p-1$, the corresponding parameter in the matrix is $\sqrt{p}$. This is manifested in the Lindblad master equation as a jump operator with an intensity of $\sqrt{p}\gamma$. Therefore, for initial states $|\Psi_\mathrm{initial}^3\rangle$ and $|\Psi_\mathrm{initial}^4\rangle$, the process of the photon number decreasing from 2 to 1 has a strength that is $\sqrt{2}$ times greater than that of the process from 1 to 0. This results in the two free photons in $|\Psi_\mathrm{initial}^3\rangle$ and $|\Psi_\mathrm{initial}^4\rangle$ being more likely to escape into the external environment at the initial stage compared to those in state $|\Psi_\mathrm{initial}^5\rangle$. In panel (d), since state $|\Psi_\mathrm{initial}^6\rangle$ is a superposition of states $|\Psi_\mathrm{initial}^3\rangle$, $|\Psi_\mathrm{initial}^4\rangle$, and $|\Psi_\mathrm{initial}^5\rangle$, the resulting probability of obtaining the hydrogen molecular cation lies between the results shown in panel (b) and panel (c). In panel (e), since there are neither photons of mode $\Omega_{12}^{\uparrow}$ nor photons of mode $\Omega_{12}^{\downarrow}$ at the initial time, the ionization of the hydrogen molecule is forbidden.
	
	\subsection{Influx process}
	\label{subsec:Influx}

\begin{figure}[htbp]
	\centerline{\includegraphics[width=.48\textwidth]{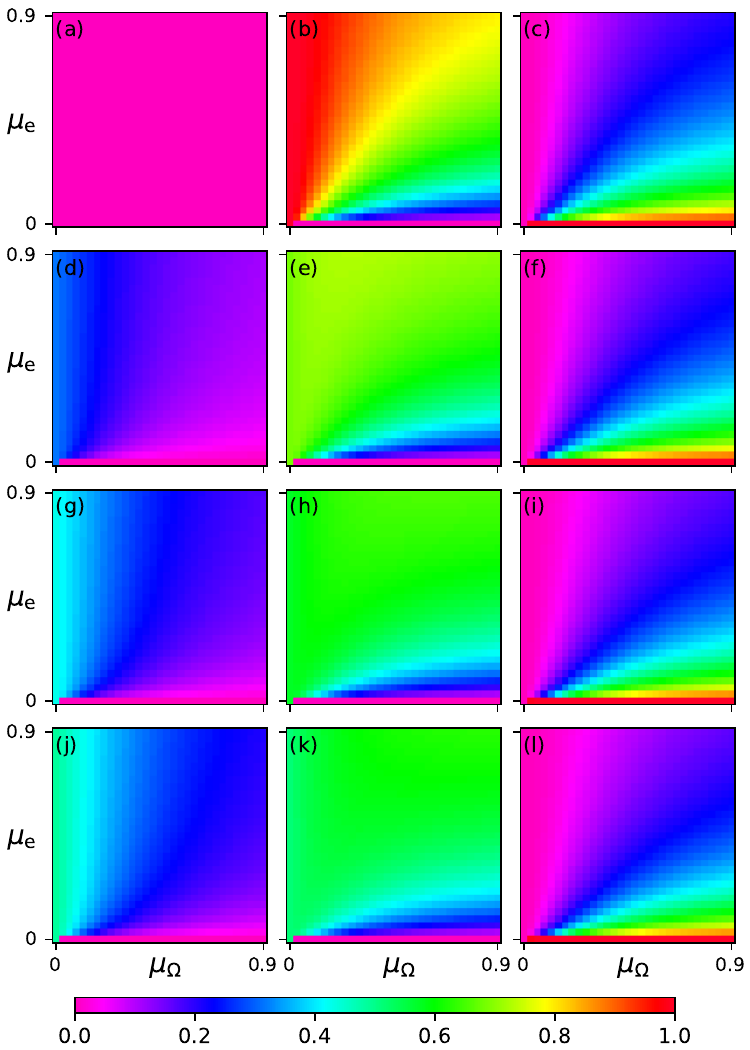}}
	\vspace*{8pt}
		\caption{(Color online) The influx effect. The picture shows the effect of different particle influx intensities on the ionization model (initial state is $|\Psi_\mathrm{initial}^6\rangle$): the horizontal and vertical axes of each panel represent the influx--dissipation ratio for photons ($\mu_\Omega$) and for electrons ($\mu_\mathrm{e}$), respectively; the first through fourth rows correspond to influx--dissipation ratio for phonons ($\mu_\omega$) of $0$, $0.3$, $0.6$, and $0.9$, respectively; the first through third columns correspond to probabilities of $|\mathrm{H},\mathrm{H}\rangle$, $|\mathrm{H}_2\rangle$, and $|\mathrm{H}_2^+\rangle$, respectively. Here, the color bar for each heatmap ranges from 0 to 1.}
		\label{fig:Influx}
	\end{figure}
	
	Having investigated the impact of dissipation on the system's evolution in Fig. \ref{fig:Model} (c), we now turn to the effects of the influx process (see Fig. \ref{fig:Influx}). First, we define the ratio of influx intensity to dissipative intensity as:
	\begin{equation}
		\label{eq:Ratio}
		\mu_k=\frac{\gamma_{k'}}{\gamma_k},
	\end{equation}
	where $\mu<1$ indicates that the overall trend in the open system is dissipative, i.e., $\gamma_{k'}<\gamma_k$. According to Eq. \eqref{eq:Ratio}, with the denominator held constant, the ratio is directly proportional to the numerator. Therefore, we can use $\mu_k$ directly to represent $\gamma_{k'}$.
	
	We now investigate the influx effect in the open system and the influence of different influx--dissipation ratios on the formation of the $\mathrm{H}_2^+$. The initial state is $|\Psi_\mathrm{initial}^6\rangle$. The intensity of all dissipation channels is uniformly set to $10^7$. Unlike in Fig. \ref{fig:Model} (c), the results in Fig. \ref{fig:Influx} show that in some cases with influx effects, energy flows not only into subspaces $|\mathrm{H}_2\rangle$ and $|\mathrm{H}_2^+\rangle$, but is also partially retained in subspace $|\mathrm{H},\mathrm{H}\rangle$. Only when $\mu_{\omega}=0$ is the energy entirely channeled into $|\mathrm{H}_2\rangle$ and $|\mathrm{H}_2^+\rangle$ (see first row of panels in Fig. \ref{fig:Influx}). Furthermore, the influx effect for photons promotes the formation of the $\mathrm{H}_2^+$, whereas the influx effect for electrons suppresses it. Specifically, the influx effect for phonons suppresses the formation of the $|\mathrm{H}_2\rangle$. However, it does not significantly promote the formation of the $|\mathrm{H}_2^+\rangle$, but rather enhances the stability of two separate hydrogen atoms ($|\mathrm{H},\mathrm{H}\rangle$). For the time-based dynamic changes of the heatmap in Fig. \ref{fig:Influx}, please refer to the \href{https://drive.google.com/drive/folders/1hZpICSbDHC9GW6X_1iPWRwpPUGFOp8mZ?usp=sharing}{animation-2.mp4}.
	
	\subsection{Model with anode} 
	\label{subsec:Anode}
	
\begin{figure*}[htbp]
	\centerline{\includegraphics[width=1.\textwidth]{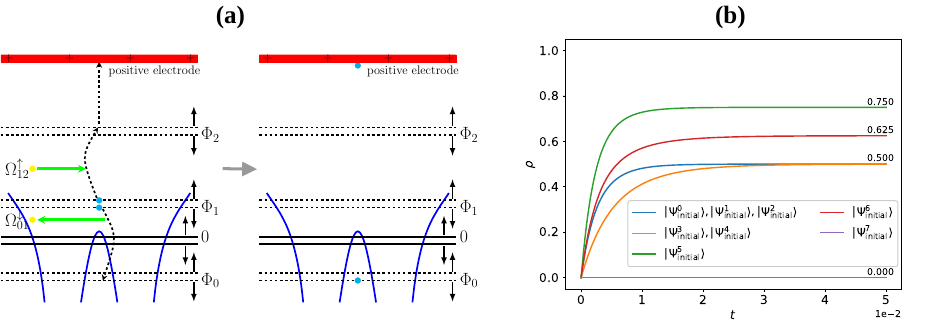}}
	\vspace*{8pt}
		\caption{(Color online) The model with an anode. In panel (a), the hydrogen molecule is assumed to be in a perfectly isolated environment, preventing free particles from dissipating to the external system. However, electrons can be absorbed by an embedded anode, which is considered as electron escape. Panel (b) shows the time-dependent probabilities of $|\mathrm{H}_2^+\rangle$ under different initial conditions.}
		\label{fig:Anode}
	\end{figure*}
	
	Fig. \ref{fig:Model} (b) presents the unitary evolution within the closed system. We now introduce a special model: an anode is incorporated into this closed system to adsorb electrons. Panel (a) in Fig. \ref{fig:Anode} shows when the electron occupies the transitional orbital, it is highly susceptible to attraction by the positive electrode, enabling it to escape from the hydrogen molecule and thereby ionize the hydrogen molecule. The anode can absorb at most one electron. This is because the transition of an electron to the $\Phi_2$ orbital can only occur when two electrons are present. When only one electron remains, the system's energy is lowered, making it less likely for the electron to escape the nuclear binding.
	
	The dissipative evolution in this new model under different initial conditions is shown in Fig. \ref{fig:Anode} (b). Since only electrons can be absorbed, while photons and phonons cannot escape, the system's energy can only flow towards subspace $|\mathrm{H}_2^+\rangle$. Consequently, we can determine the maximum probability of the hydrogen molecular cation for different initial conditions. According to Eq. \eqref{eq:InitialAtOrbital}, when atomic orbitals hybridize into molecular orbitals, the spin-up electron has an equal probability of occupying the excited orbital ($\Phi_1$) or the ground state orbital ($\Phi_0$). This implies that a spin-up electron has only a $\frac{1}{2}$ chance of transitioning to the $\Phi_2$ orbital for eventual absorption by the anode. The same logic applies to a spin-down electron. Based on this mechanism, the final probability of obtaining the hydrogen molecular cation is calculated as follows:
	\begin{itemize}
		\item For initial states $|\Psi_\mathrm{initial}^0\rangle$ to $|\Psi_\mathrm{initial}^4\rangle$, which involve only one type of free photon (where the photon number only affects the evolution speed, not the outcome), the probability is $\frac{1}{2}$.
		\item For $|\Psi_\mathrm{initial}^5\rangle$, which contains both types of free photons, enabling both spin channels, the probability is $\frac{3}{4}$.
		\item For $|\Psi_\mathrm{initial}^6\rangle$, which is a superposition of $|\Psi_\mathrm{initial}^3\rangle$, $|\Psi_\mathrm{initial}^4\rangle$, and $|\Psi_\mathrm{initial}^5\rangle$, the resulting probability is derived as $\frac{5}{8}$.
		\item For $|\Psi_\mathrm{initial}^7\rangle$, which contains no free photons, the probability is 0.
	\end{itemize}
	
	\section{Conclusion and Outlook}
	\label{sec:Conclusion}
	
	In this work, we have constructed a unified QED-Lindblad framework that systematically compares hydrogen molecule ionization across closed, dissipative, and influx-driven open systems, and revealed the distinct regulatory roles of photon and phonon dissipation — a key differentiation from previous studies. Within this framework, we have systematically investigated the quantum dynamics of hydrogen molecule ionization in closed quantum systems, as well as open quantum systems driven by dissipation and particle influx. Our key findings reveal that
	\begin{itemize}
		\item The system exhibits a strong tendency to form the stable neutral hydrogen molecule ($|\mathrm{H}_2\rangle$) under a wide range of conditions.
		\item The interplay among dissipation channels ($\gamma_{\Omega}$, $\gamma_\mathrm{e}$, $\gamma_{\omega}$) critically governs both the stabilization time and final state distribution, with $\gamma_{\Omega}$ acting as a key promoter of molecular stability and significantly accelerating the stabilization process.
		\item The introduction of particle influx ($\mu_k$) leads to more complex energy partitioning, enabling population retention in the atomic state ($|\mathrm{H},\mathrm{H}\rangle$) --- a phenomenon absent in purely dissipative dynamics.
		\item The ionization pathway shows high sensitivity to the initial quantum state, particularly the composition and number of photons, which control the accessible excitation channels for electrons of different spins.
		\item These insights are further confirmed by an anode model, showing that the maximum ionization probability is fundamentally limited by orbital hybridization and initial state configuration, reaching a theoretical maximum of $\frac{3}{4}$ under optimal conditions.
	\end{itemize}

	Looking forward, several promising directions emerge from this study. First, extending our model to include non-Markovian environments would provide a more realistic description of systems where memory effects and coherent feedback from the environment are significant. Second, investigating more complex molecular systems, such as heteronuclear diatomic molecules or those with more electrons, would test the generality of our findings and uncover new collective quantum effects. Finally, an experimental realization of this setup, for instance, using ultracold atoms in optical cavities or quantum simulator platforms, would be a critical step towards verifying our theoretical predictions and harnessing this quantum-controlled chemistry in practice.
	
\section*{Acknowledgments}
The study was funded by the China Scholarship Council, project number 202108090483. The author acknowledges the Center for Collective Use of Ultra-High-Performance Computing Resources (\url{https://parallel.ru/}) at Lomonosov Moscow State University for providing supercomputer resources that contributed to the research results.

\appendix

\onecolumngrid

	\section{List of symbols, abbreviations and notation}
	\label{appxsec:List}
	
	The list of symbols, abbreviations and notation is shown as follows:
	\begin{longtable}{ll}
    		\caption{List of symbols, abbreviations and notations.}\\
    		\label{appxtab:List}\\
    		\hline
    		\multicolumn{1}{l}{\textbf{S.A.N.}} & \multicolumn{1}{l}{\textbf{Descriptions}}\\
    		\hline
    		\endfirsthead
    
    		\multicolumn{2}{c}{{\bfseries -- continued from previous page}}\\
    		\hline
    		\multicolumn{1}{l}{\textbf{S.A.N.}} & \multicolumn{1}{l}{\textbf{Descriptions}}\\
    		\hline
    		\endhead
    
    		\hline
    		\multicolumn{1}{l}{{Continued on next page}}
    		\endfoot
    
   		\hline
    		\endlastfoot
    
    		QED & Quantum electrodynamics\\
    		QME & Quantum master equation\\
    		RWA & Rotating wave approximation\\
    		PTSIM & Precise time step integration method\\
    		$\eta$ & Dimensionless coupling parameter ($\eta=\max(\frac{g}{\hbar\omega_\mathrm{c}}, \frac{g}{\hbar\omega_\mathrm{a}})$)\\
    		$\omega_\mathrm{a}$ & Atomic frequency\\
    		$\omega_\mathrm{c}$ & Cavity frequency\\
    		$\uparrow$, $\downarrow$ & Spin-up and spin-down\\
    		$\mathrm{e}^\uparrow$ & Electron with spin-up\\
    		$\mathrm{e}^\downarrow$ & Electrons with spin-down\\
    		$\Phi_0$ & Bonding molecular orbital (ground state)\\
    		$\Phi_1$ & Antibonding molecular orbital (excited state)\\
    		$\Phi_2$ & Transitional orbital (unstable, far from nuclei)\\
    		$0_1$, $0_2$ & Atomic orbitals of two hydrogen atoms\\
    		$p_1$, $p_2$, $p_3$, $p_4$, $p_5$ & Photon numbers for respective modes\\
    		$L$ & Covalent bond state: $L = 0$ (formed), $L = 1$ (broken)\\
    		$k$ & \parbox{9cm}{Nuclear state: $k = 0$ (together in one cavity), $k = 1$ (scattered in different cavities)}\\
    		$l_1$, $l_2$ & Electron states: $0$ ($\Phi_0$), $1$ ($\Phi_1$), $2$ ($\Phi_2$), $\varnothing$ (free electron)\\
    		$\Omega_{12}^{\uparrow}$ & Photon mode for $\uparrow$ electron transition $\Phi_1 \to \Phi_2$\\
    		$\Omega_{12}^{\downarrow}$ & Photon mode for $\downarrow$ electron transition $\Phi_1 \to \Phi_2$\\
    		$\Omega_{01}^{\uparrow}$ & Photon mode for $\uparrow$ electron transition $\Phi_0 \to \Phi_1$\\
    		$\Omega_{01}^{\downarrow}$ & Photon mode for $\downarrow$ electron transition $\Phi_0 \to \Phi_1$\\
    		$\omega$ & Phonon mode frequency\\
   		$\hat{\rho}$ & Density operator\\
    		$\hat{H}$ & Hamiltonian operator\\
    		$\hat{L}$ & Lindblad superoperator\\
    		$\hat{L}_k$, $\hat{L}_{k'}$ & Lindblad operators for dissipation and influx channels\\
    		$\hat{A}_k$ & \parbox{9cm}{Jump operator for dissipation, with Hermitian conjugate $\hat{A}_k^{\dag}$}\\
    		$\hat{a}$, $\hat{a}^{\dag}$ & Annihilation and creation operators for photons/phonons\\
    		$\hat{\sigma}$, $\hat{\sigma}^{\dag}$ & Relaxation and excitation operators for electrons\\
    		$\hbar$ & Reduced Planck constant ($h/2\pi$)\\
    		$g$ & Coupling strength\\
    		$g_{\Omega_{01}^{\uparrow}}, g_{\Omega_{01}^{\downarrow}}$ & Coupling strengths for $\Omega_{01}$ photon modes\\
    		$g_{\Omega_{12}^{\uparrow}}, g_{\Omega_{12}^{\downarrow}}$ & Coupling strengths for $\Omega_{12}$ photon modes\\
    		$g_{\omega}$ & Coupling strength for covalent bond formation/breaking\\
    		$\zeta$ & Tunneling strength\\
    		$\gamma_k$ & Dissipation rate for $k \in \mathcal{K}$ (particle escape)\\
    		$\mathcal{K}$ & Set of dissipation channels\\
    		$\gamma_\Omega$ & Photon dissipation strength\\
    		$\gamma_\mathrm{e}$ & Electron dissipation strength\\
    		$\gamma_\omega$ & Phonon dissipation strength\\
    		$\gamma_{k'}$ & Influx rate for $k' \in \mathcal{K}'$ (particle inflow)\\
    		$\mathcal{K}'$ & Set of influx channels\\
    		$\mu_k$ & Influx--dissipation ratio ($\mu_k = \gamma_{k'}/\gamma_k$)\\
    		$\mu_\Omega$ & Influx--dissipation ratio for photons\\
    		$\mu_\mathrm{e}$ & Influx--dissipation ratio for electrons\\
    		$\mu_\omega$ & Influx--dissipation ratio for phonons\\
    		$\Delta t$ & Time step\\
    		$\varepsilon$ & Scaled time step ($\Delta t / 2^M$)\\
    		$M$ & Number of scaling steps (usually $M = 20$)\\
    		$I$ & Identity matrix\\
    		$A$ & General matrix (for matrix exponential)\\
    		$T_{a,0}$ & Truncated Taylor series (up to 4th order)\\
    		$T_{a,n}$ & Recursively computed matrix for $n \ge 1$\\
    		$|\Psi_{\mathrm{initial}}^i\rangle$ & Initial states for $i = 0, 1, \ldots, 7$\\
    		$|\Psi_\mathrm{f}\rangle$ & Fixed part of the initial state (remaining qubits)\\
    		$|\mathrm{H},\mathrm{H}\rangle$ & \parbox{9cm}{Subspace of two independent hydrogen atoms ($L=1$, no covalent bond)}\\
    		$|\mathrm{H}_2\rangle$ & \parbox{9cm}{Subspace of neutral hydrogen molecule ($L=0$, $k=0$, both electrons present)}\\
    		$|\mathrm{H}_2^+\rangle$ & \parbox{9cm}{Subspace of hydrogen molecular cation ($L=0$, $k=0$, exactly one electron)}\\
    		$\mathrm{span}\{\cdots\}$ & Linear span of basis states\\
    		$\mathrm{AND}$ & Logical AND operator (both conditions true)\\
    		$\mathrm{XOR}$ & Logical exclusive OR operator (exactly one condition true)\\
    		$t_\mathrm{stb}$ & \parbox{9cm}{Stabilization time (combined probability of $|\mathrm{H}_2\rangle$ and $|\mathrm{H}_2^+\rangle$ exceeds 0.999)}\\
    		$t_\mathrm{stb,max}$ & Maximum stabilization time for given $\gamma_\omega$\\
    		$t_\mathrm{stb,min}$ & Minimum stabilization time for given $\gamma_\omega$\\
    		$t_{\mathrm{mol}}$ & Stabilization time of neutral hydrogen molecule $|\mathrm{H}_2\rangle$\\
    		$t_{\mathrm{ion}}$ & Stabilization time of hydrogen molecular cation $|\mathrm{H}_2^+\rangle$\\
    		$10^{-5}$ & Convergence threshold for $t_{\mathrm{mol}}$ and $t_{\mathrm{ion}}$\\
	\end{longtable}

\twocolumngrid

\bibliography{bibliography}
	
\end{document}